\documentclass[10pt, conference, compsocconf]{IEEEtran}

\ifCLASSINFOpdf

\else

\fi

  \usepackage[nocompress]{cite}
\usepackage{algorithm}
\usepackage{multirow}
\usepackage{graphicx}
\usepackage{algorithm}
\usepackage{algorithmic}
\usepackage{array}
\usepackage{subfigure}
\usepackage{amsthm}

\usepackage[justification=centering]{caption}
\usepackage{url}
\usepackage{cite}
\usepackage{setspace}
\usepackage{amsmath,amssymb,amsfonts}
\usepackage{algorithmic}
\usepackage{graphicx}
\usepackage{textcomp}
\usepackage{enumerate}
\usepackage{amsfonts}
\usepackage{amsmath} 
\usepackage{amssymb}
\usepackage{verbatim}
\usepackage{color}
\usepackage{float}
\usepackage{setspace}
\usepackage{threeparttable}

\usepackage{cite}

\begin{document}

\title{V2V-Based Task Offloading and Resource Allocation in Vehicular Edge Computing Networks\\ 
}

\author{Junjin He\IEEEauthorrefmark{2}, ~
	Yujie Wang\IEEEauthorrefmark{2}, 
	Xin Du\IEEEauthorrefmark{2},
	Zhihui Lu\IEEEauthorrefmark{2}\IEEEauthorrefmark{1},~

	\\
	\IEEEauthorblockA{\IEEEauthorrefmark{2}\textit{School of Computer Science,}
		\textit{Fudan University,}
		Shanghai, China 
	}

}

\maketitle
	
\begin{abstract}
	In the research and application of vehicle ad hoc networks (VANETs), it is often assumed that vehicles obtain cloud computing services by accessing to roadside units (RSUs). However, due to the problems of insufficient construction quantity, limited communication range and overload of calculation load of roadside units, the calculation mode relying only on vehicle to roadside units is difficult to deal with complex and changeable calculation tasks. In this paper, when the roadside unit is missing, the vehicle mobile unit is regarded as a natural edge computing node to make full use of the excess computing power of mobile vehicles and perform the offloading task of surrounding mobile vehicles in time. In this paper, the OPFTO framework is designed, an improved task allocation algorithm HGSA is proposed, and the pre-filtering process is designed with full consideration of the moving characteristics of vehicles. In addition, vehicle simulation experiments show that the proposed strategy has the advantages of low delay and high accuracy compared with other task scheduling strategies, which provides a reference scheme for the construction of Urban Intelligent Transportation in the future.
\end{abstract}



\maketitle

\section{Introduction}
\label{sec:introduction}
\PARstart{W}{ith} the development of smart cities, intelligent applications have been integrated into all aspects of people's lives, such as transportation, production, medical care, education, entertainment, etc.\cite{b1}. As the core of the city, transportation is the most important part of the implementation of smart cities. In order to promote the reform of the transportation system, intelligent transportation systems (ITS) have emerged. ITS aims to provide better services for drivers and passengers, including traffic accident analysis, traffic flow prediction, driving route planning, traffic light management and autonomous driving assistance \cite{b2}-\cite{b4}. However, such advanced services are usually computation-intensive tasks and couldn't be completed in a limited time by the vehicle alone. Therefore, offloading tasks to the facilities with powerful computing resources can effectively improve computing efficiency, reduce experiments, and ensure service quality\cite{b5,b6}.

In order to support the the exchange of information between vehicles, infrastructures and public networks, the Vehicular Ad-Hoc Networks (VANETs) \cite{b7,b8} emerged as the most important communication technology in ITS. VANETs mainly comprise vehicle-to-vehicle (V2V) and vehicle-to-infrastructure (V2I) communications mode. Based on the above two wireless local area network technologies, vehicles can contact with other vehicles and surrounding facilities even access the core network at any time, which give the chance of offloading tasks to other infrastructures and Center Could (CC). In traditional centralized cloud computing scenarios, all tasks are offloaded to remote places and  computed centrally in large data centers. Although centralized computing has inexhaustible computing resources, it will spend a lot of time on data transmission because it is too far away from the vehicle.

In recent years, Mobile Edge Computing (MEC) \cite{b9} technology has been widely studied as a new computing paradigm. The deployment of servers is gradually migrating from remote cloud to the network edge, which provides users with lower latency and higher reliability services. In VANET, RSUs are such edge servers built on the roadside, which allows dedicated short-range communication (DSRC) with other on-board unit (OBU) mounted on vehicle. In related studies, many researchers chose to offload and execute tasks on RSUs\cite{b10}--\cite{b19}. After tasks are completes on RSUs, the results are returned back to the vehicle. Although offloading tasks to the RSU can greatly shorten the delay, there are still some issues with completely relying on RSU. First of all, it is impossible to make RSU cover all urban roads, not only due to geographical and environmental constraints, but also because of the high construction cost of RSU. Secondly, even in places covered by RSU, when a large number of vehicles gather together and a large number of the tasks need to be executed, the RSUs will face the peak of offloading requests, which far exceeds the resources that RSU can provide. 

Considering computing resources of a large number of vehicles are wasted. If vehicles are used as communication and computing infrastructure \cite{b20} and these rich resources are integrated, the quality of services and applications will be greatly improved. Multiple vehicles that can communication with each other compose a vehicular cloud \cite{b21}. It enables vehicles to offload tasks to surrounding vehicles. Tasks will be distributed to different vehicles, so that the computing resource of each vehicle in this cloud can be utilized. Moreover, these vehicles form a distributed structure, and tasks can be executed by different in parallel. In recent years, some researchers have also shifted their focus to vehicle-to-vehicle offloading\cite{b22}--\cite{b27}. Tasks are classified into dependent tasks and independent tasks according to whether they need to be executed in a fixed order. Some authors proposed new algorithms based on game theory, graph theory or reinforcement learning to optimize offloading strategy, and evaluated their methods through delay, energy consumption, resource utilization, etc. It has been proved that V2V offloading mode indeed have contributed to better resource utilization and higher computation performance.

In this paper, we focus on V2V task offloading mode, regarding vehicles as the main computing resource providers. We design an online pre-filtering task offloading system (OPFTO) to make full use of vehicle idle computing resources. At the same time, we also prove the necessity of considering vehicle mobility in offloading, and put forward the corresponding solutions. 

The main contributions of this paper are as follows:

1) We propose OPFTO, a task offloading system that tasks are offloaded between vehicles and vehicles. The system mainly includes two stages: selecting candidate vehicle and allocating tasks to vehicles. Accurate filtering can effectively improve the success rate of offloading, and an excellent offloading strategy can also feedback to promote more accurate filtering parameters.

2) We design a HGSA algorithm for the task offloading process. The algorithm is based on the improvement of simulated annealing algorithm, and can quickly get a nearly optimal offloading strategy. HGSA can effectively shorten the task completion time and realize the rational allocation of resources.

3) In particular, we consider the impact of vehicle mobility. In order to improve the accuracy of offloading strategy, we designed a pre-filtering process to avoid the task failure due to the sudden departure of vehicles. The process will automatically adjust the parameters in the pre-filtering process according to the historical data to adapt to the latest situation.

The remainder of this paper is organized as follows. In Section II, we describe the related work about task offloading. The system model and problem formulation are introduced in Section III. The structure of OPFTO system and corresponding task offloading algorithm are presented with details in Section IV. Section V shows the evaluation experiment and numerical results. Finally, the paper is concluded in Section VI.

\section{Related Work}
In this section, we review the state of the art for task offloading in vehicular network. In general, researches can be classified into two categories: V2R/V2I task offloading and V2V task offloading.

\subsection{V2R/V2I Task Offloading}
In order to obtain more powerful computing resources, most of papers tend to offload tasks to RSUs or CC. Since RSU is located at the edge of the network and close to the vehicles, it has almost become the first choice for offloading. Over the past decade, most articles have investigated and discussed the offloading strategies based on V2R and V2I.

Some studies design offloading strategies based on game theory \cite{b12,b13,b14}. \cite{b16} designed a payoff function and constructed a distributed optimal response algorithm for computing offloading game. It is proved that the offloading probability of each vehicle can converge to a unique equilibrium under certain conditions. C. Shu et al. \cite{b30} studied the offloading of dependent tasks to edge servers. Considering the dependency between subtasks and the competition among multiple edge servers, they proposed a new offloading scheme based on game theory to reduce the overall completion time of applications.

Some other studies are based on reinforcement learning \cite{b13,b18,b25}. H. Ke et al. \cite{b13} designed a heterogeneous vehicle network MEC system considering the changes of channel state and available bandwidth. They propose an adaptive computational offloading method based on deep reinforcement learning (ACORL). ACORL can intelligently learn policy to solve the trade-off between energy consumption, bandwidth allocation and execution delay. Y. Liu et al. \cite{b18} regard the vehicles as mobile edge servers (VES), combined with fixed edge servers (FES), to provide computing services for users. They expressed task offloading and resource allocation as Markov processes, and proposed two offloading strategies based on Q-Learning and deep reinforcement learning respectively. Their scheme achieves better performance than pure VES or FES methods.

Other papers take different approaches. P. Dai et al. \cite{b12} studied the task allocation to MEC, and they proposed a multi arm bandit learning algorithm. Through online learning of real-time workload distribution, a utility table is established to determine the best solution, and it is updated according to the feedback signal of task allocation. N. VanDung et al. \cite{b14} studies the tasks offloading to heterogeneous RSUs. They propose an algorithm design of joint contention window control and offloading decision formulated as a mixed-integer non-linear problem to maximize system utility. J. Sun et al. \cite{b17} Studied the influence of the execution order of tasks on the offloading decision. They designed a hybrid intelligent optimization algorithm to maximize the offloading utility of the system and obtain the approximate optimal solution with low time complexity.

In addition, there are also papers devoted to the impact of vehicle mobility. H. Vu huy et al. \cite{b15} jointly analyzed the handover, multiple rates and backhaul costs, and designed a model to analyze and predict the cost. Their mode can reduce the cost by 17\% compared with the baseline.

\subsection{V2V Task Offloading}
Although RSUs and CC can provide high computing power, they still have some disadvantages. For example, they need to bear additional infrastructure deployment costs. Fortunately, there are a huge number of vehicles in cities today. Vehicles can also provide us with rich computing resources. In recent year, some papers began to turn their focus on from to the V2V offloading strategies.

M. LiWang et al. \cite{b22} studied the allocation problem of V2V communication in vehicular cloud environment. They proposed a randomized graph job allocation mechanism via hierarchical tree-based subgraph isomorphism with low complexity. Their method has been proved to have better performance than the baseline method when considering the job completion time and data exchange cost.

J. Zhao et al. \cite{b23} proposed a collaboration method based on MEC and cloud computing. Tasks can be offloaded not only to MEC servers, but also to vehicles. Based on game theory, they designed a collaborative computation offloading and resource allocation optimization (CCORAO) scheme. This scheme can effectively improve the system utilization and calculation time.

J. Zhang et al. \cite{b24} studied task offloading in V2V and V2I communication modes. Based on the load balancing of the computation resources among the MEC servers, they proposed an approximate computing offloading scheme, namely ALBOA. This scheme can make more effective use of the computing resources of the edge server and reduce the processing delay.

Q. Qi et al. \cite{b25} studied the problem of offloading services to vehicle computing nodes. They proposed a knowledge driven (KD) IoV service offloading decision framework, which provides the optimal policy directly from the environment. Deep reinforcement learning is used to obtain the optimal solution, and continuous online learning is carried out during vehicle service execution, so as to adapt to environmental changes.

Y. Sun et al. \cite{b26} regard vehicles as computing nodes. Tasks can be offloaded not only directly to nearby vehicles, but also to remote vehicles with the help of RSU. Authors proposed an adaptive learning-based task offloading (ALTO) algorithm to minimize the average offloading delay. At the same time, they modified the existing multi-armed bandit (MAB) algorithm to input sensing and event sensing, so that ALTO algorithm can adapt to the dynamic vehicle task offloading environment. Compared with the classical upper confidence limit algorithm, the average delay of this algorithm is reduced by 30\%.

G. Qiao et al. \cite{b27} regard the vehicle as an edge computing resource and offload the tasks to the vehicles. They propose a collaborative task offloading and output transmission mechanism to guarantee low latency as well as the application-level performance.

\section{Modeling}
In this section, we first overview the whole architecture for task offloading in vehicular network. Then we analyze the time overhead required for the communication and computing process, respectively. Finally, the problem is transformed into an optimization problem, and the objective function and constraints are defined.

\subsection{Overview}
In this paper, we mainly study how to make full use of the idle computing resources of surrounding vehicles through vehicular cloud. As shown in Figure 1, RSUs and CC are only responsible for collecting and processing the data from sensors, and providing necessary information for vehicles when needed, such as the location, speed and destination. In the vehicular cloud, each vehicle is assigned different numbers and types of tasks according to their location and computing power. In order to focus on the research of V2V offloading, in this scenario, tasks only flow in the vehicular cloud and will not flow to the upper RSUs and cloud.

We assume that there are some vehicles are driving in the same direction on the highway. Let $V=\{v_1,v_2,...,v_n\}$ denote the set of vehicles and the number of vehicles is n. For each vehicle vi, it can be represented by a tuple $\{f,p,s\}$, in which f represents the computing power of the vehicle ( CPU cycles/s ), $p$ represents the current position of the vehicle, and $s$ represents the real-time speed of the vehicle. At the beginning, these vehicles can communicate to each other and form a vehicular cloud. 

At this time, the k-th vehicle is having some tasks to be processed, so we call it the task vehicle. Let $Q=\{q_{1},q_{2},\ldots,q_{m}\}$denote the set of tasks and the number of tasks is $m$. For each task $q$, it can be presented by a tuple $\{S, C, D_{limit}\}$, where $S$ is the amount of data of the task, $C$ is the computational complexity of the task (CPU cycles/bit), $D_{limit}$ is the maximum allowable delay of the task.

These compute-intensive tasks are classified into four categories: image processing, video processing, interactive game and augmented reality, and their computational complexity id denoted by $C_A,C_B,C_C,C_D$. Since all tasks are compute-intensive, it is obvious that a single vehicle can’t complete these tasks in a limited time. The task vehicle needs to divide the task into multiple subtasks, and then distribute them to different vehicles for parallel execution. There is no dependency between the two subtasks. All tasks are split into the same-size subtasks, and the number of subtasks can be calculated as
\begin{equation}m'=\sum_{i=1}^{m} \frac{S_{i}}{e}.\label{eq1}\end{equation}
where $e$ is the minimum unit of the subtask.

For each independent subtask, it will be offloaded to any of the other vehicles or be executed locally. In other word, each subtask has $n$ choices. The offloading strategy of all subtasks can be expressed as a matrix with size $m' \times n$ as
\begin{equation}X=\left[ \begin{matrix}
x_{11} & x_{12} & \dots & x_{1n} \\
x_{21}& x_{22} & \dots & x_{2n} \\
\dots \\
x_{m'1} & x_{m'2} & \dots & x_{m'n} \\
\end{matrix} \right].\label{eq2}\end{equation}
where $x_{ij}=1$ denotes that the i-th subtask will be offloaded to the vehicle  $v_j$, otherwise, $x_{ij}=0$. Specially, if $x_{ik}=1$, it means that the i-th will be execute on local.

For each subtask, it can only be assigned to one vehicle, unless the subtask cannot be completed on the original vehicle, a new vehicle will be assigned to it. As for vehicles, each one can be assigned multiple tasks. Once it completes a subtask, it will immediately return the results to the task vehicle.

The entire offloading can be divided into three stages. First, the task vehicle offloading all or part of the subtasks to other vehicles. Then, all subtasks will be executed on their assigned vehicles. Finally, the computing result will be return to the task vehicles. A task can be considered completed only when the results of all subtasks of this task are returned to the task vehicle. A summary of the main notations used in this paper is shown in Table 1.

\begin{table}
	\caption{Summary of Main Notations}
	\label{tab1}
	\setlength{\tabcolsep}{3pt}
	\begin{tabular}{|p{50pt}|p{180pt}|}
		\hline
		Notation& 
		Description \\
		\hline
		$V$& 
		The set of vehicles \\
		$n$&
		The number of vehicles \\
		$f_{i}$&
		The computing capability of i-th vehicle \\
		$p$&
		The position of i-th vehicle \\
		$s$&
		The speed of i-th vehicle \\
		$k$&
		The task vehicle \\
		$Q$&
		The set of tasks \\
		$m$&
		The number of tasks \\
		$S$&
		The data size of the task \\
		$C$&
		The computational complexity of the task (cycles/bit) \\
		$D_{limit}$&
		The delay constraint \\
		$e$&
		The minimum size of subtask \\
		$m'$&
		The number of subtasks \\
		$x_{ij}$&
		The offloading strategy \\
		$r_{direct}$&
		The transmission rate between two neighbor vehicles \\
		$\mu$&
		The number of hops between two vehicles \\
		$R_{ij}$&
		The transmission rate between two vehicles \\
		$D_{tran}$&
		The transmission delay \\
		$D_{comp}^{loc}$&
		The computing delay \\
		$D_{comp}^{V2V}$&
		The computing delay of subtask executed on \\
		$sub(i)$&
		The task to which i-th subtask belongs \\
		$D_{ij}$&
		The completion time of subtask \\
		$T_{j}$&
		The total time of a vehicle \\
		$T_{total}$&
		Total completion time of all tasks \\
		$T_{stay}$&
		Total time of all tasks \\
		\hline
	\end{tabular}
\end{table}

\subsection{Communications Model}

In the vehicular cloud, the task vehicle can obtain the basic information of all vehicles in the cloud from RSUs, including computing power, location and speed. Therefore, the task vehicle can calculate the Euclidean distance between vehicles according to their position. When the Euclidean distance between two vehicles is within the communication range, two vehicles can communicate directly. Otherwise, they need to communicate through multi hop.

When subtasks are to be offloaded to other vehicles for execution, the task vehicle first need to transmit the data of the subtasks to other vehicles through V2V mode. According to Shannon formula [40], the transmission rate (MB/s) between the vehicle-k and its neighbor vehicle-i can be calculated as\begin{equation}r=\frac{B}{m'-1} log_{2}(1+\frac{p \cdot h}{N}).\label{eq3}\end{equation}where, $B$ represents the channel bandwidth, $p$ represents the channel power, and $h$ means the channel gain and $N$ represents the noise power. Since the vehicle transmits tasks outward at the same time, the channel bandwidth will be divided into $m'-1$ segments.

However, not all vehicles in the cloud can communicate with each other directly. When the Euclidean distance between two vehicles exceeds the communication range of the vehicle, they can communicate with each other through multi hop. Let $\mu$ denotes the minimum hops between two vehicles. Therefore, the transmission rate between vehicles through multi hop communication can be calculated as\begin{equation}R=\frac{1}{\sum_{\mu} \frac{1}{r}\label{eq4}}.\end{equation}

It can be seen that when the power of difference vehicles is similar, the total transmission power can be approximately expressed as \begin{equation}R=\frac{B}{\mu(m'-1)} log_{2}(1+\frac{p \cdot h}{N}).\label{eq5}\end{equation}

After that, the transmission delay when the subtask is offloaded can be obtained by\begin{equation}D_{tran}=\frac{e\mu(m'-1)}{B\cdot log_{2}(1+\frac{p \cdot h}{N})}.\label{eq6}\end{equation}

\subsection{Computing Model}
After transmitting a subtask to another vehicle or leaving it locally, the next step is to compute it.

When the subtask is executed locally, its computing delay depends on the computing capacity of the task vehicle. Then, the computing delay of the i-th subtask executed locally is calculated as\begin{equation}D_{comp}^{loc}=\frac{C_{i} \cdot e}{f_{loc}}\quad , i=1,2,\ldots,m'.\label{eq7}\end{equation}
where $C_{i}$ is the computational complexity of the i-th subtask and $f_{loc}$ is the computing capacity of the task vehicle

When other vehicles have idle computing resources, subtasks will be offloaded to other vehicles for execution. The calculation delay of the i-th subtask execute on the j-th vehicle is calculated as \begin{equation}D_{comp}^{V2V}=\frac{C_{i}  \cdot e}{f_{j}}\quad , i=1,2,\ldots,m'.\label{eq8}\end{equation}

When the subtask is successfully executed, the result is returned to the task vehicle through the vehicle. For computing intensive tasks, the size of calculation result is far less than the input data, so the return time of the result is far less than the transmission delay and calculation delay, which can be ignored. Therefore, we express the completion time of the task as 
\begin{equation}D_{ij}=\left\{ 
\begin{aligned}
\frac{C_{i}  \cdot e}{f_{j}} \quad ,j=k \\
\frac{C_{i}  \cdot e}{f_{j}}+\frac{e}{R} \quad ,j \neq k \\
\end{aligned}
\right.\label{eq9}\end{equation}
where, $j = k$ represents the subtask executed locally, and the completion time is the calculation delay. When j is not equal to k, the subtask is offloaded to other vehicles, and the completion time is the sum of transmission delay and calculation delay.

As described above, each subtask will be executed on their assigned vehicle. Each subtask will take a different time from being transmitted, executed to returning results. Only when all subtasks of a task are completed, the task can be regarded as completed. Therefore, the completion time of a task can be calculated as\begin{equation}D_{task}=max\{\sum_{i}^{q} x_{ij} \cdot D_{ij}\}\quad , j=1,2,\ldots,n.\label{eq10}\end{equation} where, the subtask with the largest completion time is the last subtask returned.

At the same time, we also need to ensure that each task is completed within its corresponding maximum tolerance delay, so we need to meet the relationship \begin{equation}max\{\sum_{i}^{q} x_{ij} \cdot D_{ij}\} \leq D_{limit}\quad , j=1,2,\ldots,n.\label{eq11}\end{equation}

Under the condition that a single task needs to meet the maximum tolerance delay, we want the overall task completion time to be as short as possible. The total task completion time depends on the vehicle the longest completion time. Therefore, we calculate the total time of each vehicle as \begin{equation}T_{j}=\sum_{i=1}^{m'} x_{ij} \cdot D_{ij} \quad , j=1,2,\ldots,n.\label{eq12}\end{equation}

All subtasks will be offloaded at the same time after the task vehicle makes the offloading strategy. Each service vehicle will return the results to the service vehicle after all subtasks on it are calculated. For task vehicles, the current task is completed only after receiving the return results of all subtasks. Therefore, the total task completion time required for offloading in V2V mode is the maximum value of the completion time required for each service vehicle, as shown in equation (13). \begin{equation}T_{total}=max\{T_{j}\}\quad , j=1,2,\ldots,n.\label{eq13}\end{equation}

In addition, due to the mobility of vehicles, each vehicle may leave the vehicular cloud at some time in the future. Therefore, it is necessary to allocate an appropriate number of tasks to the vehicle according to the time it stays in the vehicular cloud. For example, a vehicle has strong computing power, but it will soon leave the vehicular cloud. At this time, we are not willing to give up his computing power, so we prefer to unload the task onto it. However, because it will leave soon, we will not assign it after assigning a certain task, which ensures that it can complete all the assigned tasks in a limited time. \begin{equation}\sum_{i=1}^{m'} x_{ij} \cdot D_{ij} \leq T_{stay}\quad , j=1,2,\ldots,n.\label{eq14}\end{equation}

\subsection{Problem Formation}
Based on the above analysis, we model and analyze the task offloading problem from vehicle to vehicle on the highway. Given that the number of subtasks is m 'and the number of vehicles is n, we hope to obtain an excellent offloading matrix X (M' * n) under some constraints to minimize the total task completion time. The optimization objective function and constraints are \begin{equation}min\{max\{\sum_{i=1}^{m'} x_{ij} \cdot D_{ij}\}\quad , j=1,2,\ldots,n.\label{eq15}\end{equation}

\begin{equation}s.t. \left\{
\begin{aligned}
& C1: x_{i1}+x_{i2}+\cdots+x_{in}=1 \quad , i=1,2,\ldots,m' \\
& C2: x_{ij}=0\quad or \quad x_{ij}=1 \\
& C3: max\{\sum_{i}^{q} x_{ij} \cdot D_{ij}\} \leq T_{limit} \\
& C4: \sum_{i=1}^{m'} x_{ij} \cdot D_{ij} \leq T_{stay} \quad , j=1,2,\ldots,n 
\end{aligned}\right . \label{eq16}\end{equation} where constraint C1 indicates that one subtask can only be offloaded to one vehicle. C2 indicates that when subtask $i$ is offloaded to vehicle $j$, $x_{ij} = 1$, otherwise $x_{ij} = 0$. C3 is the completion time required for each task, which should be within the tolerable limit time. C4 is that for each vehicle, the sum of the completion time of all its subtasks should be less than its stay time in the vehicular cloud, so as to ensure that the vehicle can complete all subtasks before leaving the vehicular cloud.

\section{Task Offloading Strategy}
In this section, we first introduce the overall process of OPFTO system, and then discuss the details of candidate vehicle selection and task allocation algorithm HGSA in OPFTO.
\subsection{OPFTO SYSTEM}
OPTFO is an online pre-filtering task offloading system, which includes a filtering process before offloading. It can obtain the optimal offloading strategy and significantly reduce the impact of vehicle mobility on the completion delay. The structure of proposed OPFTO system is shown in Figure 2, it is divided into two steps: selecting candidate vehicles and allocating tasks to vehicles.

The first step is selecting reliable candidate vehicles for each subtask, which is designed to ensure the effectiveness of future task offloading. Due to the mobility of vehicles, they may leave the vehicular cloud before completing their assigned tasks. If a vehicle leaves the vehicular cloud without completing all the tasks, the uncompleted tasks will need to be assigned again, which will cause additional time overhead. Therefore, we hope to offload each subtask to a reliable vehicle and avoid ineffective offloading as much as possible. For a subtask, a vehicle that carries too many subtasks or has a short stay in the car cloud is more likely to cause the task not to be executed. In order to judge whether a vehicle is suitable for a subtask, we comprehensively consider three indicators, namely, the stay time of vehicle in vehicular cloud, the completion time of the subtask, and the task success rate of the vehicle. Then use the above three indicators to calculate a value to guide the judgment. Finally, each subtask owns its corresponding candidate vehicles.

In the second step, all subtasks are allocated to the appropriate vehicles. As mentioned in section III, all tasks will be divided into multiple independent subtasks that can be executed in parallel. These subtasks execute on different vehicles at the same time, which will greatly shorten the completion time of the whole task compared with sequential execution. The aim of this step is to find the most suitable vehicle for each subtask to minimize the total completion time of the all tasks. We first calculate the completion time of each subtask on its all candidate vehicles and form a matrix. Then we use our proposed offloading algorithm HGSA to calculate an excellent allocation scheme in a limited time. Finally, vehicles are allocated for subtasks according to the gained scheme. 

In addition, whenever a vehicle completes all its subtasks or leaves the vehicular cloud ahead of time, the task success rate of the vehicle will be updated. If the task success rate increases, the vehicle will get more offloading opportunities. If the task success rate decreases, it indicates that the risk of offloading to the vehicle increases. The two-way feedback can improve the the effectiveness of offloading decision and deduce the total completion delay.

\subsection{Candidate Vehicles}

Selecting candidate vehicles for each subtask is important, because it can avoid the additional time overhead caused by vehicle mobility. For subjective or objective reasons, when a vehicle leaves the original vehicular, it will not be able to communicate with the task vehicle. The uncompleted subtasks on this vehicle need to find a new vehicle for computing, which is a time-consuming process. However, if the subtask knows that the vehicle will leave and adds this vehicle to the blacklist, it can plan ahead and reduce the delay. To demonstrate the necessity of this step, Figure 3 shows the effect of existence of candidates on the overall task completion delay.

Given four subtasks of the same size and four vehicles with different computing powers. The time required for these vehicles to complete a subtask is [5,5,3,4]. At time 0, all subtasks are offloaded, and then at time 4, vehicle 4 will leave the vehicle cloud. Figure (a) (b) shows the offloading situation before and after the vehicle leaves without candidate processing. At the beginning, because the subtask did not know that vehicle 4 would leave, it chose the optimal offloading strategy. In this way, the minimum time required to complete the four tasks is 5. However, vehicle 4 suddenly leaves the vehicular cloud when the time is nearly to 4, resulting in the result can't be returned to task vehicle, so the failed subtask is offloaded to the vehicle 3 for execution. In this case, the minimum time required to complete the four tasks is 7. As shown in Figure (c), if it is known from the beginning that vehicle 4 will leave early in the future, the subtask will not initially use vehicle 4 as a candidate vehicle, and the subtask will achieve optimal offloading among the other three vehicles. The time delay required in this way is 6, which is better than the case without candidate processing.

In order to correctly judge whether a vehicle should be the candidate for subtasks, we comprehensively considered the following three factors:
\begin{itemize}
	\item \emph{The completion time of the task:} According to the analysis in Section III, we use equation (9) to calculate the completion time of each subtask on each vehicle.
	\item \emph{The stay time of vehicle:} The stay time refers to the time from the beginning to the moment the vehicle leaves the vehicular cloud. Since our hypothetical scene is a highway, we use the freeway mobility model \cite{b28} to mimic the behavior of the vehicles. Then the trajectory of the vehicles in a longer time range can be predicted and the stay time can be estimated.
	\item \emph{History offloading success rate:} Assuming that the number of i-th vehicle's historical participation in task offloading is $y_i$, and the number of tasks successfully completed by the vehicle is $x_i$, the history offloading success rate of this vehicle can be calculated by $\frac{x_i}{y_i}$.
\end{itemize}
Among them, the relationship between task completion time and vehicle stay time reflects the objective fact of whether the task can be successfully completed on the vehicle, and the historical offloading success rate reflects the possibility of the vehicle makes task failed due to subjective reasons.

Then, we used the historical data of the above three indicators to train a mode based on machine learning, which outputs the parameter $\lambda$ to indicate the suitability of adding a vehicle to the candidate vehicle. The training set format is shown in Table 2.

\begin{table}
	\caption{The training set for predicting the suitability of adding a vehicle to the candidate vehicle}
	\label{tab2}
	\setlength{\tabcolsep}{3pt}
	\begin{tabular}{|p{15pt}|p{30pt}|p{50pt}|p{60pt}|p{55pt}|}
		\hline
		Task& 
		stay time &
		completion time &
		history success rate &
		success / failure \\
		\hline
		1 & 30 & 20 & 0.85 & success \\ 
		2 & 25 & 15 & 0.80 & failure \\
		3 & 28 & 17 & 0.75 & failure \\
		$\cdots$ & $\cdots$ & $\cdots$ & $\cdots$ & $\cdots$\\
		500 & 34 & 23 & 0.46 & success \\
		\hline
	\end{tabular}
\end{table}

We obtain $\lambda$ by inputting the stay time, task completion time and history offloading success rate of the vehicle. The value of $\lambda$ is between [0,1]. When $\lambda=1$, the vehicle must be added to the candidate vehicle. When $\lambda < 1$, the vehicle is added to the candidate vehicle with the probability of $\lambda$.  For example, when $\lambda= 0.8$, it means that the task has 80\% chance of being completed on this vehicle, so we also have a 80\% probability to add the vehicle to join the candidate vehicle for the task.

At the same time, this model is also an online model. With the progress of time, whenever a new task is completed or failed, it will update the indicator of "history offloading success rate" in the model and carry out new training for the model.

\subsection{HGSA Algorithm}

After selecting candidate vehicles for each task, the next step is to find the optimal task allocation strategy that minimizes the overall task completion delay.

Before assigning tasks to vehicles, each subtask needs to calculate the completion time on its candidate vehicle and form a cost-time matrix $C$. This matrix has $m'$ rows and $n$ columns, representing $m'$ subtasks and $n$ vehicles. In matrix $C$, each element is different due to the different computing and communication capabilities of vehicles and different computational complexity of subtasks. According to the candidate mechanism discussed in the previous section, some vehicles will not be added into certain subtasks' candidates due to their early leavings. For those vehicles, we set the completion time of the subtask executed on them as infinity. As for other vehicles that can finish the subtask normally, their completion time is calculated according to equation (8). In this way, the cost time matrix of all subtasks and all vehicles is formed. 

Next, our goal is to allocate $m'$ independent subtasks to $n$ vehicles, let them execute on different vehicles in parallel, and then find the minimum total task completion time. This problem can be regarded as a 0-1 integer linear programming problem, which is proved to be a NP hard problem. For this problem, we can use greedy algorithm to find the optimal solution of the problem. However, with the increase of the number of subtasks, the complexity of using greedy algorithm increases greatly. In the face of more subtasks, greedy algorithm can not give us the best results in effective time.

In this work, we propose an half green simulated annealing (HGSA) algorithm which allocates the most appropriate vehicles for subtasks to minimize the task completion delay, as shown in Algorithm 1. This algorithm combines the advantages of greedy algorithm and simulated annealing, which is able to gain the optimal strategy in a limited time. In this algorithm, the input includes the number of subtasks $m'$, the number of vehicles $n$, the cost-time matrix $C$ records the completion time of each subtask on its candidate vehicles calculated by equation (8), and parameters used to control the annealing iterative process of $T_{max}$, $T_{min}$, $\alpha$. The output of the algorithm contains the optimal task offloading matrix $X$ with the size of $m' \times n$, in which $x_{ij}=1$ represents the subtask i is offloaded to vehicle j. Another output parameter is vehicle total time vector $\vec{\tau}$ with size of $n$, it indicates the total time for each vehicle to complete all subtasks on it.

\begin{algorithm}[htb]
	\caption{HGSA Algorithm}
	\label{alg1}
	\begin{algorithmic}[1]
		\REQUIRE Number of subtasks $m'$; Number of vehicles $n$; Cost-time matrix $C$; Initial temperature $T_{max}$; Terminal temperature $T_{min}$; Cooling rate $\alpha$
		\ENSURE Task offloading matrix $X$, Vehicle total time vector $\vec{\tau}$
		\STATE initial $X=0$, $\vec{\tau}=0$
		\FOR{$i=1$ to $m'$}
		\STATE $j^*\gets min\{\tau_j+C_{ij}\},j=1,2,\ldots,n$
		\STATE Check constraints of the equation (16)
		\STATE $x_{ij^*}=1$, $\tau_j=\tau_j+C_{ij}$
		\ENDFOR 
		\STATE $f(X)=max\{\vec{\tau}\}$, according to the equation (13)
		\STATE $T= T_{max}$, $X_{old}=X$, $f(X)_{old}=f(X)$
		\WHILE{$T < T_{min}$} 
		\STATE $X_{old}=X$, $f(X)_{old}=f(X)$
		\STATE Randomly exchange two subtasks with different types, get $X_{new}$ and $\vec{\tau}_{new}$
		\STATE Check constraints of the equation (16)
		\STATE $f(X)_{new}=max\{\vec{\tau}_{new}\}$, according to the equation (13)
		\STATE $\Delta = f(X)_{new}-f(X)_{old}$
		\IF{$\Delta < 0$} 
		\STATE $X=X_{new}$ and $\vec{\tau}=\vec{\tau}_{new}$
		\ELSE 
		\STATE $p=\exp(\Delta/T)$ 
		\ENDIF 
		\IF{$ramdom.p < p$} 
		\STATE $X=X_{new}$ and $\vec{\tau}=\vec{\tau}_{new}$
		\ENDIF 
		\STATE $T=T \cdot \alpha$ 
		\ENDWHILE
		\RETURN $X$ and $\vec{\tau}$
	\end{algorithmic} 
\end{algorithm}

As shown in the steps 1-7 of HGSA algorithm, it is mainly to find a feasible solution of task offloading matrix $C$. We firstly initialize the matrix $C$ and $\vec{\tau}$ to 0, and then find an available vehicle from the candidates for each subtask one by one. In order to obtain the optimal result in a short time, we didn't randomly assign subtasks to vehicles in this phase, but sought a sub-optimal solution as much as possible. With the principle of greed, we try to ensure the overall delay is minimized in every offloading operation. For each subtask, we calculate the vehicle total time after offloading it to each candidate vehicle and then select the smallest one in step 3. Then check the constraints such as whether the total time of this vehicle exceeds the stay time, if the minimum vehicle meets the constraint, select it as the target vehicle for offloading, just like the step4, step5. It is worth noting that we did not pay too much attention to the order of subtasks, but to repeat the experiment many times so as to eliminate the effect of order. If the order of the subtasks is to be considered, it is necessary to find the smallest element from the entire matrix $X$, rather than from $n$ vehicles. This time cost is $m'$ times that of our algorithm. As the number of subtasks increases, the result is not available in a limited time. After each subtask is assigned to the vehicle, the offloading matrix $X$ and the total time $\tau$ of the corresponding vehicle are updated. Until all tasks perform the above operations, the maximum total time in vehicles is the total completion delay $f(X)$ of the all subtasks.

After initialization, we will perform simulated annealing. Similar to the annealing process of metal, starting from the initial temperature $T_{max}$, with the continuous decline of temperature parameter $T$, combined with the probability jump characteristics, randomly find the global optimal solution of the objective function in the solution space until the temperature is lower than the termination temperature $T_{min}$, and end the annealing operation. In each iteration from step 9 to step 24, we first need to generate a new strategy by randomly exchanging the subtasks with different types and check whether the new strategy meets the constraints.If the restrictions are met, calculate the corresponding $f(X)$ according to equation (13), so as to obtain the total delay required to complete all tasks (i.e. the maximum of the cumulative delay of each vehicle). Next in step 14, compare the value of the total completion delay $f(X)$ of the new and old strategy. If $f(X)_{new}$ is smaller, it means the new strategy is better and we will accept the new strategy $X$ and update vehicle total time $\vec{\tau}$. When $f(X)_{new}$ is large, it may be a case of jumping out of the local optimum. Therefore, the new strategy is accepted according to a certain probability, which is related to the current temperature. Determine the probability according to $p=\exp(\Delta/T)$. Finally, the current temperature is multiplied by the cooling factor to obtain the new temperature. The above operation is a complete iteration, and then the current temperature is multiplied by the cooling coefficient to obtain the new temperature to enter the next iteration. Until the temperature drops to the lowest temperature, the iteration ends and return the optimal strategy $X$ and vehicle total time $\vec{\tau}$. Based on above analysis, HGSA algorithm can obtain an offloading strategy that guarantees the optimal delay in a short time.

\section{Experiment and Evaluation}
In this section, we first introduce the experimental environment, datasets and parameter settings. Then, we evaluate the performance of our proposed algorithm by comparing with the baseline algorithm. Finally, we discuss the significance of candidate process in OPFTO.

\subsection{Simulation Setup}
We simulated our experiment environment in MATLAB by implementing a three-lane highway of length of 10km and some moving vehicles. The trajectory of vehicles is selected from the Madrid trace \cite{b29}, which is a set of synthetic trace containing real information from three highways (A6, M40 and M30) in Madrid, Spain. Each vehicle has different computing ranging from $4\times10^{6}$ to $2\times10^{7}$. The tasks are classified into 4 categories, and their computational complexity is 30,15,40,20 and each task can be divided into 50-100 same subtasks. The main simulation parameters are summarized in Table 3.

\begin{table}
	\caption{Summary of Simulation Parameters}
	\label{tab3}
	\setlength{\tabcolsep}{3pt}
	\begin{tabular}{|p{160pt}|p{70pt}|}
		\hline
		Parameters& 
		Value \\
		\hline
		Number of vehicles & 
		5-50 \\
		Range of V2V communication &
		150 m \\
		Size of subtask &
		$10^{6}$ \\
		Vehicle computing power &
		$4\times10^{6}$-$2\times10^{7}$ \\
		Vehicle transmission power &
		1.3 w \\
		Gaussian white noise power &
		$3\times10^{-13}$ \\
		Channel gain &
		4 \\
		V2V Link bandwidth &
		2 MHz \\
		Data size of vehicular task &
		100 kb \\
		Task computational complexity&
		50 cycles/bit\\
		Speed of vehicles &
		60-100 km/h \\
		Number of subtasks&
		50-150 \\
		Task computational complexity&
		30, 15, 40, 20\\
		
		\hline
	\end{tabular}
\end{table}

In order to simulate the real movement of the vehicles, we selected the track records of 20 vehicles on A6 highway within 30 minutes and set the communication range of vehicle to 150 m. At the beginning, these vehicles are all in the vehicular cloud and can communicate with each other, and their locations and connectivity are shown in the Figure 4.

\subsection{Performance Comparison}
For the performance comparison, we consider four other algorithms as follows:

Random: Each subtask will be randomly assigned to any vehicle that the task vehicle can connect to.

Distance First (DF): All vehicles are sorted according to the distance between themselves and task vehicle from near to far. Then the subtasks are offloaded to the vehicle according to the above sequence and the stay time of the vehicle.

Strongest Computing power First (SCF): All vehicles that can provide computing resources are sorted  by their computing power. And then the tasks are offloaded to these vehicles according to the sorting results and the longest task execution time that can be provided.

Shortest Stay-time First (SSF): Subtasks are preferentially assigned to the earliest departing vehicle  with the shortest stay-time.

\subsubsection{Impact of Vehicle Number}
In order to study the influence of the number of vehicles on the delay, we fixed the number of subtasks to 400, and then set the number of vehicles to 10 to 20.

Figure 5 shows the relationship between the average delay of completing a subtask and the number of vehicles. It can be seen that with the increase of vehicles, the average delay of subtasks of the five algorithms shows an upward trend. However, HGSA, SSF and RANDOM algorithms have shorter average completion delay than SCF and DF. This is because SCF algorithm focuses on computing power and DF algorithm focuses on communication power. Both of them consider only one part of the task completion delay.

Figure 6 shows the change of total task delay as the number of vehicles in the vehicular cloud increases. It can be seen that the total delay obtained by the five offloading algorithms will decrease with the increase of the number of tasks in the train set. However, with the increase of the number of vehicles, the effect of SCF and DF algorithms is very poor and remains unchanged. This is because although SCF algorithm and DF always unload tasks to the same vehicle, this weakens the benefits of parallel execution of subtasks on different vehicles. In addition, although HGSA has similar performance with SSF and random in average offloading delay, HGSA performs better in total task completion delay. This is because our HGSA algorithm makes full use of the resources of other vehicles and makes use of more parallel advantages. In conclusion, HGSA algorithm is better than the other four algorithms. Compared with SSF, the total delay is reduced by 54\%, and compared with random, the total delay is reduced by 19\%.

\subsubsection{Impact of Task Number}
Next, we discuss the impact of the number of tasks on the delay when the number of vehicles is 20.

Figure 7 shows the average delay of subtasks as the number increases. We can find that the four algorithms have some similar trends. It can be seen that compared with other algorithms, HGSA algorithm always has the least delay regardless of the number of subtasks. Especially when the number of subtasks is small, the advantages of HGSA algorithm are more obvious. This is because when the number of subtasks is small, the priority offloading of other algorithms to a vehicle will lead to more idle vehicles, while HGSA balances the computing resources of all vehicles to obtain the shortest delay.

Figure 8 shows the total completion delay as the number of subtasks increases. It can be seen from the figure that HGSA and RANDOM increase linearly with the number. SCF and DF first rise rapidly and then remain unchanged, while SSF first remains unchanged and then rises rapidly. This is because the total completion delay depends on the vehicle with the longest completion time in the vehicle. When the number of tasks is large enough, SCF, DF and SSF all need more vehicles to join the offloading. They inevitably select a vehicle with a long stay time, resulting in a large total completion delay. HGSA will not offload tasks to some but all vehicles, so that more tasks can run in parallel.


\subsubsection{Impact of Stay time}
We select vehicle-A, vehicle-B and vehicle-C from 20 vehicles as task vehicles, representing three different scenarios in which other vehicles stay in the vehicular cloud for a short, medium and long time. The stay time of other vehicles in the vehicular cloud is as Figure 9. In which, vehicle-A means that the surrounding vehicles will leave the vehicular cloud soon, and the vehicular cloud with vehicle-C as the task vehicle will last longer, that is, the vehicles in it can have a longer completion time.

Figure 10 shows the total completion delay under different length of residence time. It can be seen that DF, SSF and SCF differ greatly under different stay time. Because the stay time is determined by the location of the vehicle, it shows that these three algorithms are greatly affected by the location of the task vehicle in the vehicular cloud. The HGSA algorithm proposed by us is very stable no matter where task vehicles are located, so our algorithm has strong robustness to the location of task vehicles. The difference of stay time distribution of other vehicles will not affect the performance of HGSA algorithm.

\subsection{Effect of Vehicular Mobility}
In our proposed OPFTO system architecture, it is very necessary to select candidate vehicles for each subtask. In order to demonstrate the importance of this step, we compared the time required for each vehicle under the HGSA algorithm with candidate process and the HGSA algorithm without candidate process (HGSA-NoCandicate). As shown in Figure 11.

We set the number of subtasks to 1400 and the number of vehicles to 20. As can be seen from the figure, when we use HGSA without selecting candicate to gain offloading strategy, all vehicles will have a similar time cost. However, in this case, some vehicles's stay-time less than the time that they need to complete all subtasks, such as 3, 4, 6, 8, 12-20. It means that they will leave the vehicular cloud before completing all subtasks and it needs more time to offload these subtasks again.

When the step of selecting candidate vehicles is added, the offloading strategy will first meet the time for vehicles to complete all subtasks within the stay time, and then make the completion time of other vehicles as close as possible. The results show that the algorithm with candidate steps not only has low delay and ensures the success rate of the task.

\section{Conclusion}
In this paper, we proposed an online strategy for task offloading and resource allocation through vehicle-to-vehicle mode. Our offloading system OPFTO includes online candidate vehicle selection and task to vehicle allocation, which can greatly reduce the overall task completion delay and prevent the execution of the task from being affected by the mobility of the vehicle. In the selection of candidate, we considered completion time of task, the stay time of vehicles and history offloading success rate to guide the judgement. Then we designed the HGSA algorithm to solve the allocation problem, which is an improved algorithm based on simulated annealing. The simulation results showed that under different parameter configurations, the HGSA algorithm performs better than other baseline schemes in terms of time delay, and our algorithm can indeed effectively avoid the additional overhead caused by vehicle mobility. In the future work, we will study the tasks with mutual dependencies and take the execution priority of subtasks into account.

\end{document}